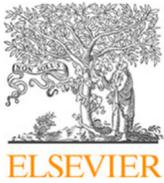
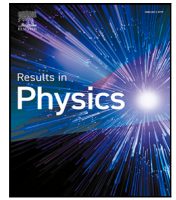

# Multi focus acoustic field generation using Dammann gratings for phased array transducers

Tatsuki Fushimi [a,b,*], Yusuke Koroyasu [c]

[a] *Institute of Library, Information and Media Science, University of Tsukuba, Kasuga 1-2, Tsukuba, Ibaraki, 305-8550, Japan*
[b] *R&D Center for Digital Nature, University of Tsukuba, Kasuga 1-2, Tsukuba, Ibaraki, 305-8550, Japan*
[c] *Graduate School of Comprehensive Human Sciences, University of Tsukuba, Kasuga 1-2, Tsukuba, Ibaraki, 305-8550, Japan*

ABSTRACT

Phased array transducers can shape acoustic fields for versatile manipulation; however, generating multiple focal points typically involves complex optimization. This study demonstrates that Dammann gratings – binary phase gratings originally used in optics to generate equal-intensity spot arrays – can be adapted for acoustics to create multiple equal-strength focal points with a phased array transducer. The transducer elements were assigned phases of 0 or $\pi$, based on a Dammann grating defined by its transition points. Simulations show that simple gratings with two transition points can generate fields with up to 12 focal points of nearly equal acoustic pressures. Compared to conventional multi-focus phase optimization techniques, the Dammann grating approach offers computational efficiency and facile reconfiguration of the focal pattern by adjusting the grating hologram. We tested this approach in numerical simulations with a hypothetical high-resolution array, achieving up to 12 focal points, and validated the efficacy of the Dammann grating in a conventional 16x16 transducer array through both simulations and experiments. This comparison highlights that while Dammann gratings effectively generate multi-focus fields, the recreation ability of these gratings in a conventional array shows a lower resolution than the hypothetical array. This study underlines the potential of adapting binary phase functions from photonics to enhance ultrasound-based acoustic manipulation for tasks requiring parallel actuation at multiple points.

## Introduction

Acoustic radiation force has emerged as a powerful tool for remotely manipulating and controlling small particles in diverse fields. Unlike other remote manipulation techniques, such as photophoretic or electrostatic forces, acoustic force offers a distinct advantage due to its ability to exert force on target objects irrespective of their material properties. There have been significant advances in the field of acoustophoresis, driven by the development of phased array transducers (PAT) in recent years [1–4]. Specifically, the dynamic multi-focal capabilities of PAT have garnered considerable interest due to their potential in enabling parallelization [5,6]. The ability to simultaneously focus on multiple targets can enhance performance and efficiency in applications such as experiment automation [7,8], acoustophoretic displays [9–11], and ultrasonic haptic displays [5,12]. One of the most straightforward methods for generating multiple high-pressure points involves using a standing wave field, as illustrated in Fig. 1. A standing wave can be generated by the superposition of two counter-propagating waves, typically generated by a transducer and reflector or by another set of transducers. Although this approach is conceptually simple, it necessitates counter-propagating waves, which may not always be practical due to spatial constraints or accessibility limitations.

An alternative approach involves specifying the desired focal position and using a phase retrieval algorithm (or acoustic hologram optimizers, Fig. 1) to determine the appropriate phase sets that produce the desired field. A wide range of techniques have been developed to realize multi-focal fields, including Eigensolver approach [5,13], iterative backpropagation (IBP) [6], GS-PAT [13], and Diff-PAT [14,15]. Although, these algorithms can be efficient, they require computational resources as they must be optimized numerically.

In this study, we introduce the use of acoustic lenses based on Dammann gratings to generate multi-focal fields. Dammann gratings [16] are binary phase gratings that produce one- or two-dimensional arrays of equal intensity in optics and have been widely used in Fourier optics. However, to date, their implementation for acoustics has remained largely unexplored. A notable advantage of the proposed method is that the multi-focal field can be directly specified by the

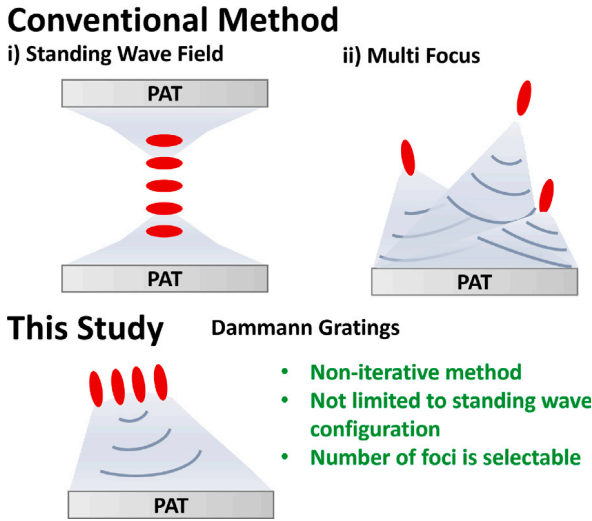

**Fig. 1.** Comparison of acoustic field generation techniques for creating multiple high pressure regions. Conventional methods: (a) standing wave fields easily create multiple high-pressure regions but require two opposing waves, (b) phase optimization methods realize custom fields but require optimization. Proposed method: (c) the Dammann grating phase profile assigns 0 or $\pi$ phases to elements, and generates multiple focus.

Dammann grating function. This eliminates the need for optimization algorithms and expands the method's applicability to a range of applications. It should be noted that petal beams [17,18] can also generate multi focus fields, yet Dammann grating presents an alternate strategy. The expansion of methodological diversity benefits the study of acoustic hologram. Moreover, where petal beams create a focused field around the propagation axis: Dammann grating forms a standing wave-like field along propagation axis.

**Methods**

Dammann grating can be defined as follows [16,19]:

$$g(x, \mathbf{d}) = \sum_{n=0}^{N} (-1)^n \text{rect}\left[\frac{x - 0.5(d_{n+1} + d_n)}{d_{n+1} - d_n}\right], \quad (1)$$

where $\mathbf{d}$ denotes a vector in a format $\mathbf{d} = [d_0, d_1, \ldots, d_{N+1}]$ and contains $N - 1$ number of transition points (the vector must be in ascending order, $d_0$ and $d_{N+1}$ must be 0 and 0.5, respectively). In this study, we explore two transition points (i.e. maximum $N = 3$), and $d_1$ and $d_2$ denote the first and second transition points, respectively. Furthermore, the rectangular function is defined as rect$(x) = 1$ if $|x| < 0.5$, 0 if $|x| \geq 0.5$. The function $g(x, \mathbf{d})$ returns a binary output ($-1$ or 1). The coordinate $x$ is a normalized source position with 100 equally spaced points spanning [0, 0.5] (to create high resolution Dammann grating grid from which the transducer array can be interpreted from). The obtained function $g(x, \mathbf{d})$ is replicated multiple times to generate a square matrix, $G_x(x, y) = [g(x), g(x), \ldots, g(x)]$, and same sets of square matrix, $G_y(x, y)$, is also created in $y$ axis by repeating the process using the same transition points in the $y$ axis. The Dammann grating for the whole array is identified by; $H(x, y) = G_x(x, y) + G_y(x, y)$. In $H(x, y)$, the locations with 0, $-2$, and 2 are replaced with values 1, $-1$, and $-1$, respectively (Post-processing method based on[1]). Then, the nearest point interpolation function is applied to $H(x, y)$ to identify the phase at the closest point to the transducer in normalized form. Finally, $+1$ and $-1$ are replaced by 0 and $\pi$, respectively to create the Dammann lens ($\phi_{\text{Dammann}}$). A visual aid for the generation process is available in the supplementary, and codes to replicate the process will be in the Data Availability section.

---

[1] https://github.com/aakhtemostafa/SSPIM.

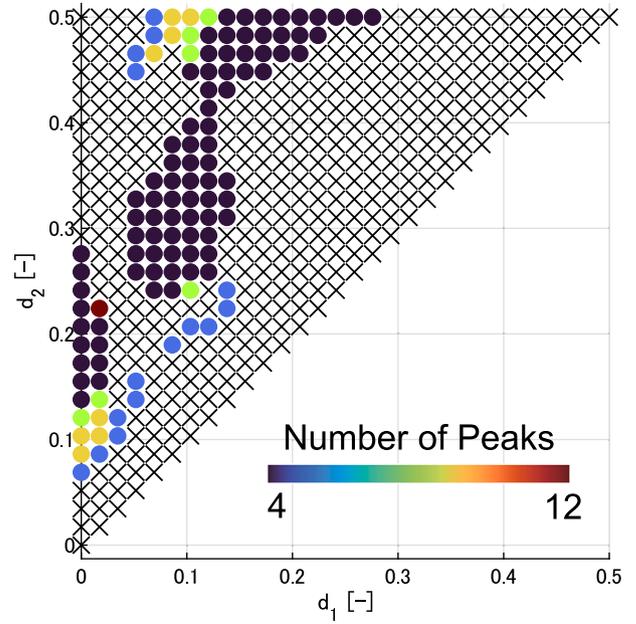

**Fig. 2.** Combination of transition points that yields "valid" multi focus fields. The color coding shows the number of focal points or peaks. Valid fields have focal points with amplitude >32.5% of the single focus and within $-3$ dB ($0.707 p_{max}$) of the highest focal point. (For interpretation of the references to color in this figure legend, the reader is referred to the web version of this article.)

**Results**

*Hypothetical high resolution array (Numerical)*

To determine the ultimate capability of Dammann gratings in the context of mid-air acoustics, we first consider an ideal case where the resolution of phased array transducer is high (81 by 81 transducers with 2-mm pitch, 40 kHz) with two transition points ($N = 2$). The pressure field generated by the Dammann gratings is calculated using the Huygens' linear superposition method ($p(\mathbf{x}, \mathbf{x_t}) = |\Sigma_t^T p_t(\mathbf{x}, \mathbf{x_t})|$), where

$$p_t(\mathbf{x}, \mathbf{x_t}, \mathbf{x_f}) = \frac{P_A}{R(\mathbf{x}, \mathbf{x_t})} e^{j(kR(\mathbf{x}, \mathbf{x_t}) + \phi)} \quad (2)$$

where $P_A = 1$, $k = \frac{2\pi f_0}{c_0}$, $\mathbf{x}$ and $\mathbf{x_t}$ denote the field and transducer position, respectively. $R(\mathbf{x}, \mathbf{x_t}) = \|\mathbf{x} - \mathbf{x_t}\|$ is the distance between the field and transducer position. Furthermore, $\phi_{\text{focal}}(\mathbf{x_t}, \mathbf{x_f}) = -\left(\frac{2\pi f_0}{c_0}\right)[d_{tf}(\mathbf{x_t}, \mathbf{x_f}) - \|\mathbf{x_f}\|]$ is the acoustic hologram for a single focus, and $d_{tf}(\mathbf{x_t}, \mathbf{x_f}) = \|\mathbf{x_t} - \mathbf{x_f}\|$, where $\mathbf{x_f}$ is the focal point. For detailed graphical scheme for the geometrical coordinate system, please refer to supplementary material. Additionally, $f_0$ and $c_0 = 346$ ms$^{-1}$ denote the acoustic frequency and speed of sound in mid-air ($\lambda = 0.00865$m), respectively. $\phi$ denotes the phase delay, and for a simple combination of Dammann gratings and focal points, $\phi = \phi_{\text{Dammann}} + \phi_{\text{focal}}(\mathbf{x_t}, \mathbf{x_f})$, and for combinations with trapping signatures, $\phi = \phi_{\text{Dammann}} + \phi_{\text{focal}}(\mathbf{x_t}, \mathbf{x_f}) + \phi_{\text{trap}}$. For a twin trap, when the transducer position is $x < 0$, $\phi_{\text{trap}} = 0$, and when $x \geq 0$, $\phi_{\text{trap}} = \pi$. For a vortex trap, based on transducer position $x$ and $y$; $\phi_{\text{trap}} = \arctan\left(\frac{y}{x}\right)$.

The acoustic pressure fields generated by Dammann gratings, with two transition points ($d_1$ and $d_2$) incremented in 30 equally spaced points between 0 and 0.5, are shown in the supplementary material. The focal point was fixed at (0, 0, 0.1) m, and the depicted pressure field is at $z = 0.1$ m. The pressure amplitude is normalized to the acoustic pressure amplitude ($p_{max}$) at the focal point when a single focus is specified. Given that $d_1 \leq d_2$, only the upper half of the combination matrix is applicable. An examination of each combination reveals that

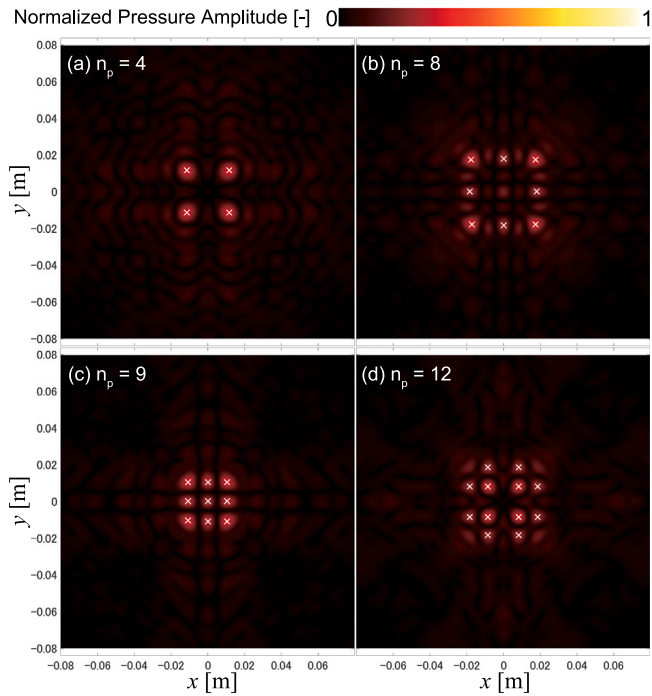

Fig. 3. Optimal multi-focus gratings that satisfy the set criteria. The transition points are shown as $(d_1, d_2)$, where $d_1$ and $d_2$ denote the normalized source positions. (a) Four focal points with transition points $(0.138, 0.379)$, (b) 8 focal points with $(0.103, 0.241)$, (c) 9 focal points with $(0.0862, 0.483)$, and (d) 12 focal points with $(0.0172, 0.224)$.

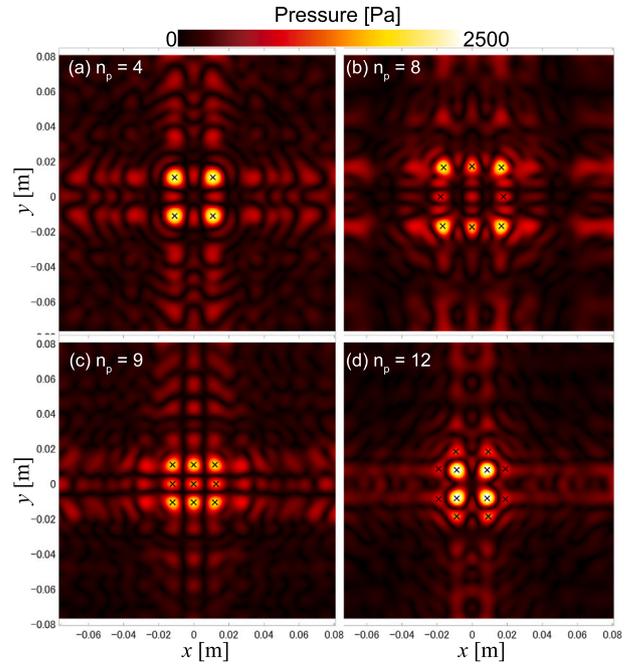

Fig. 4. Dammann grating applied to a 16 × 16 phased array transducer. The configurations are the same as Fig. 3: (a) 4 focal points, (b) 8 focal points, (c) 9 focal points, and (d) 12 focal points. The focal spots are less defined due to the lower resolution of the phased array.

Dammann gratings can specify a wide range of acoustic pressure fields, and adjusting $d_1$ or $d_2$ can express wide range of acoustic pressure field, as shown in Fig. 2 and Supplementary Figure S1.

To identify useful Dammann gratings, criteria defining desired multi-focal field properties must be established. A useful multi-focal field is considered to have multiple focal spots with significant acoustic pressure and focal spots of nearly equal acoustic pressure strength. A peak finding algorithm was utilized to identify focal spots in the 2D acoustic pressure fields (x–y plane at z = 0.1 m) generated by the Dammann gratings. The algorithm initially detects all local maxima in the 2D field matrix, including those in flat regions. The local maxima are then sorted in descending order of acoustic pressure magnitude. The algorithm iteratively discards any local maxima that is within 5 mm of a higher pressure local maxima, retaining only the local maxima that are sufficiently separated and exhibit the highest acoustic pressures.

Focus spot groups are considered to produce a "valid" multi-focal field when the maximum acoustic pressure at a focal spot ($p^{peak}$) is greater than 32.5% of the single focus pressure ($0.325 p_{max}$). Additionally, the acoustic pressures at the focal spots are in the range between $0.707 p^{peak}$ and $p^{peak}$ (within −3 dB). These criteria ensure that the focal spots have significant acoustic pressures and are of nearly equal acoustic pressure strength, producing a useful multi-focal field. It is important to acknowledge that the 32.5% threshold for valid focal points was empirically determined and, as such, is inherently arbitrary. While this value serves as a practical benchmark for identifying high-pressure points in multi-focal fields in the context of this study, it is not intended to be a universal standard. Researchers are encouraged to adjust this threshold based on their specific experimental conditions or objectives. Although the choice of this particular threshold does not influence the main contributions of this paper – which focus on the method of generating and analyzing multi-focal fields rather than on the threshold itself – future studies may further investigate the effects of varying threshold values. Such investigations could lead to the establishment of more standardized criteria for defining valid focal points across different applications

The combinations of transition points that yielded valid focal spots are summarized in Fig. 2. The majority of combinations were considered invalid (75.3%), but valid multi-focal fields with 4, 5, 8, 9 and 12 peaks (number of focal points, $n_p$) were obtained. The percentages of fields with 4, 5, 8, 9, and 12 peaks were 18.3%, 3.23%, 1.29%, 1.72%, and 0.215%, respectively. Based on these groups, one grating with the highest pressure amplitude, $p^{peak}$, was selected for each number of focal spots, $n_p$. This narrowed down the selection to one grating for each $n_p$. The results for $n_p = 4, 8, 9$, and 12 are shown in Fig. 3, where the white crosses indicate the locations of the identified focal points (see the supplementary material for $n_p = 5$). The phase profiles and 3D visualization of acoustic field for each grating are shown in the Supplementary Material. Although the field with $n_p = 4$ and 12 (Fig. 3(a)–(d)) appear similar to each other, this is inevitable due to the relatively simple filtering process. The gratings with a large number of focal points are of interesting as they can potentially serve as "array-generating" multi-focal fields for experimental automation in biology, chemistry, and medicine, effectively replacing standing wave fields [8].

*Conventional 16 × 16 array (Numerical and experimental)*

Until now, an ideal PAT with high spatial resolution is assumed, but the identified traps should translate well into a conventional phased array. Specifically, a 16 by 16 phased array with Murata MA40S4S transducer (40 kHz, 10-mm diameter, $p_0 = 0.221$ Pa m V$^{-1}$ with 20 V [8]) was assumed. A directivity function ($D(\theta) = \frac{2J_1(kr\sin\theta)}{kr\sin\theta}$) was added to simulate piston-source transducers and the results are as shown in Fig. 4. In principle, the acoustic trap, as simulated in Fig. 3(a)–(d), are well-recovered in PAT. However, it is less focused and has more ambient noise than ideal case. In particular, the difference in the pressure amplitude between peaks are evident in Fig. 4(b) and (d), as shown in the supplementary material. We further note that $n_p = 5$ in the supplementary material does not fully recover the same pressure distribution as that in Fig. 4. However, despite these challenges; the conventional PAT is sufficiently resolved to generate Dammann gratings.

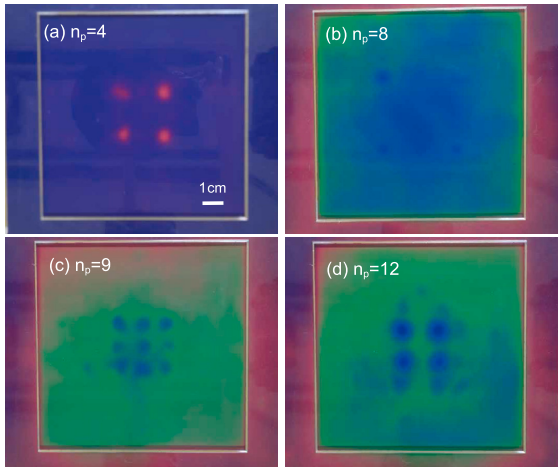

**Fig. 5.** Experimental results using a thermochromic sheet to replicate the numerical simulation. Due to varying acoustic pressure levels, the visibility peaks at different time steps. Since the field remained consistent throughout the experiment, the image was captured at the point of maximum visibility (camera triggered simultaneously with the array activation, where $t_c$ is the time elapsed since activation). The configurations are the same as in Fig. 3: (a) 4 focal points ($t_c = 40$ s), (b) 8 focal points ($t_c = 24$ s), (c) 9 focal points ($t_c = 16$ s), and (d) 12 focal points ($t_c = 57$ s).

We experimentally replicated the simulated conditions shown in Fig. 4 using the SonicSurface platform, developed by Rafael et al. [20]. The array consists of 256 Murata MA40S4S transducers, consistent with the configuration used in our numerical simulations. A 20 V supply from an external DC power source was applied to fully match the conditions of the simulation. At the focal plane ($z = 100$ mm), we placed a thermochromic sheet (20–32 °C, C-TASK) to visualize the focal points. It is well established that regions of high acoustic pressure can induce thermal effects, making this a common method for confirming the acoustic pressure field [21,22]. Recently, such thermal effects have also been used to quantify the pressure field [23]. Using this visualization method, we confirmed that the experimentally obtained pressure fields closely align with the simulated results, as illustrated in Fig. 5 (see supplementary video for raw footage). The Dammann gratings that successfully translated from the high-resolution hypothetical array to the conventional array (i.e., $n_p = 4$ and 9) exhibited clear multi-focal point fields, as predicted by the numerical simulations. However, those that did not translate well showed poor field patterns. These experimental results validate the feasibility of generating Dammann gratings with conventional arrays, but also highlight the limitations imposed by the array's spatial resolution, as predicted in both the simulations and experimental findings.

**Discussion**

*Translative and rotational property*

One of the distinguishing advantage of the Dammann grating, when compared to the hologram optimization method, is the capability to translate the multi-focus with the change in single-focus hologram, and does not require re-optimization of the field. This is due to the fact that Dammann grating is a binary phase hologram with 0 and $\pi$, and it creates phase singularities [1]. The translation capability of the multi-focus lens using $n_p = 4$ is as shown in Fig. 6(a)–(b), and the mean pressure amplitude at the peaks were 1948, 2260, and 2686 Pa at $-5\lambda$, $-2.5\lambda$, and 0 shift, respectively. The field can also be rotated by rotating the Dammann grating phase (using "imrotate" function in MATLAB, the applied phase is shown in the supplementary material). The field rotations with $\frac{\pi}{8}$, $\frac{\pi}{4}$ radians are shown in Fig. 6(c)–(d), respectively. The mean pressure amplitude at the focus stays relatively constant with pressure amplitude of 2209 and 2531 Pa. These characteristics exhibit a similar simplicity to that of the trap lens (such as twin, vortex, bottle trap lens), particularly in determining suitable acoustic holograms for linear translation and rotation of acoustic fields using Dammann grating. Acoustic focus can also be shifted vertically while retaining the multi-focusing ability. However, increasing the focal distance results in greater separation distances between each focal point (refer to the supplementary material for detailed analysis).

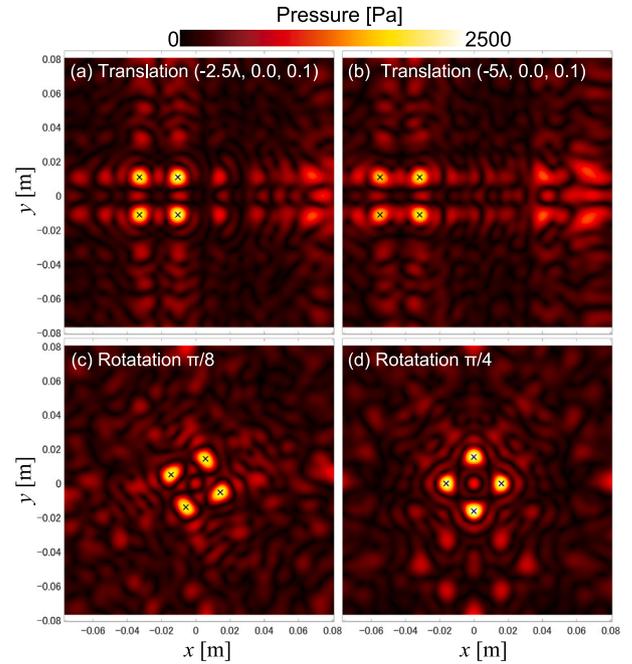

**Fig. 6.** Translation and rotation of the four focal points ($n_p = 4$) with a 16 × 16 phased array transducer. (a)–(b) show horizontal translation of the focal points by shifting the focal position by $-2.5\lambda$ and $-5\lambda$, respectively. (c)–(d) show rotation of the Dammann grating and corresponding rotation of the focal points by $\pi/8$ and $\pi/4$ radians, respectively.

*Testing compatibility with other holograms*

A characteristics that is not confirmed to be shared between the trap signature and Dammann grating is the ability to combine the lens with the optimized lens such as IBP [6] (see supplementary). Moreover, we numerically show that it is possible to combine trap signatures such as twin and vortex trap and can double the number of focal points, or apply orbital angular momentum around the focal point; further enhancing the utility and flexibility of Dammann gratings in practical scenarios. However, there is a limitation that the spatial resolution is insufficient even in an 81 by 81 grid, limiting applicable configurations to low multi-focal points (such as $n_p = 4$, particularly for orbital angular momentum). See the supplementary material for the details.

*Insights regarding multi-focus holograms*

When multi-focus fields are generated using phased array transducers, a rule of thumb used to set "sensible" target acoustic pressure amplitude involves assuming that the sum of acoustic pressure amplitude at each focus should not exceed the single focus pressure amplitude $p_{max}$ (i.e., each peak shares the acoustic pressure amplitude from the single focus amplitude) [14,15]. Thus, for a conventional multi-focus optimizer, the target normalized pressure amplitude for multi-focus with the same pressure amplitudes are the reciprocal of the number of peaks ($n_p$) [14]. The examination of Fig. 3 reveals that the mean normalized pressure amplitudes are 0.325, 0.284, 0.287, 0.307, and 0.269 for $n_p = 4, 5, 8, 9,$ and 12, respectively. This demonstrates that a higher target pressure amplitude can be set than the rule of thumb. This is an important insight into the limits of PAT, and aids in designing more appropriate performance test for acoustic hologram optimizers.

### Conclusion

In this study, we investigated relatively simple Dammann gratings with two transition points. The number of transition points can be increased, potentially leading to the discovery of Dammann gratings with a larger number of focus. However, the limitation is currently imposed by the spatial resolution of the array, which constrains our ability to accurately generate and control a higher number of focal points. For the PAT, the limitation is the physical size of the transducers, and it can be improved by the application of metamaterials [24]. Application of Dammann gratings in underwater acoustics can also be envisioned. Acoustic field that can act in lieu of standing wave exhibits potentials in creating arrays for biology/chemistry [25], additive manufacturing, and medical applications [26,27] where it is difficult to generate standing waves.

In summary, this paper presents a method that uses Dammann gratings to simply and effectively generate multi-focal acoustic fields. The Dammann grating function directly specifies the multi-focal field, eliminating the need for computational optimization. The findings of the study provide new insights into multi-focal acoustic pressure field and indicate that the Dammann gratings are promising for applications requiring PAT.

### CRediT authorship contribution statement

**Tatsuki Fushimi:** Writing – review & editing, Writing – original draft, Visualization, Validation, Supervision, Software, Resources, Project administration, Methodology, Investigation, Data curation, Conceptualization. **Yusuke Koroyasu:** Writing – review & editing, Visualization, Data curation, Conceptualization.

### Declaration of competing interest

The authors declare that they have no known competing financial interests or personal relationships that could have appeared to influence the work reported in this paper.

### Acknowledgment

We gratefully acknowledge the support of AI tools, OpenAI's GPT-4, and Anthropic's Claude. The authors have diligently reviewed and verified all generated outputs to ensure their accuracy and relevance. We would like to thank Editage [http://www.editage.com] for editing and reviewing this manuscript for English language.

### Supplementary data

See supplementary material for a visual guide for the generation process of Dammann gratings, pressure field outputs for all combinations of Dammann gratings with two transition points, coordinate system, visualization and analysis of the $n_p = 5$ field, holograms and 3D visualizations of the acoustic fields for the selected gratings ($n_p = 4, 5, 8, 9, 12$), effect of changing resolution $81 \times 81$ vs $16 \times 16$, holograms for translation and rotation operations including the applied phase transformations, detailed results of attempts to combine Dammann grating with an IBP optimizer, and additional analysis on the vertical shift capability while retaining multi-focusing ability and the effect of combining Dammann gratings with trapping signatures. Please also see the supplementary videos for the experimental verification of the Dammann gratings.

Supplementary material related to this article can be found online at https://doi.org/10.1016/j.rinp.2024.108040.

### Data availability

We already made data publicly available on Github (https://github.com/DigitalNatureGroup/Dammann_Grating_Acoustics), and is also availe in Zenodo as a permanent respository (https://doi.org/10.5281/zenodo.14249227)